\documentclass[aps,prb,preprint, a4paper,amsfonts,amssymb,floats,titlepage,fleqn,lengthcheck,twocolumn,
tightenlines,12pt,showpacs,floatfix,superscriptaddress,floatfix,footinbib,hidelinks,longbibliography]{revtex4-2}%balancelastpage, citeautoscript  
\usepackage{derivative}
\usepackage[para]{threeparttable}
\usepackage[table, dvipsnames]{xcolor}
\usepackage[utf8]{inputenc}
\usepackage[T1]{fontenc}
\usepackage[version=4]{mhchem}
\usepackage{comment}
\newcommand{\Onlinecite}[1]{\hspace{-1 ex} \nocite{#1}\citenum{#1}} 
\usepackage{subfiles}
\usepackage{mathastext}
\usepackage{amssymb}
\usepackage[nice]{nicefrac}
\usepackage{amsmath}
\usepackage{amsthm}

\usepackage{eqparbox}

\usepackage{mathtools}
\setcitestyle{super}
\usepackage{xcolor}
\usepackage[normalem]{ulem}
\usepackage[normalem]{ulem}
\usepackage{nicematrix}
\usepackage{enumitem}
\usepackage{titlesec}
\usepackage{hyperref}
\titleclass{\subsubsubsection}{straight}[\subsection]
\newcounter{subsubsubsection}[subsubsection]
\renewcommand\thesubsubsubsection{\thesubsubsection.\arabic{subsubsubsection}}
\titleformat{\subsubsubsection}
  {\normalfont\normalsize\bfseries}{\thesubsubsubsection}{1em}{}
\titlespacing{\subsubsubsection}
{0pt}{3.25ex plus 1ex minus .2ex}{1.5ex plus .2ex}
\makeatletter
\renewcommand\paragraph{\@startsection{paragraph}{5}{\z@}%
  {3.25ex \@plus1ex \@minus.2ex}%
  {-1em}%
  {\normalfont\normalsize\bfseries}}
\renewcommand\subparagraph{\@startsection{subparagraph}{6}{\parindent}%
  {3.25ex \@plus1ex \@minus .2ex}%
  {-1em}%
  {\normalfont\normalsize\bfseries}}
\def\toclevel@subsubsubsection{4}
\def\toclevel@paragraph{5}
\def\toclevel@paragraph{6}
\def\l@subsubsubsection{\@dottedtocline{4}{7em}{4em}}
\def\l@paragraph{\@dottedtocline{5}{10em}{5em}}
\def\l@subparagraph{\@dottedtocline{6}{14em}{6em}}
\makeatother
\usepackage{xr}
\usepackage{hyperref}
\usepackage{xr}
\usepackage{xr-hyper}
\usepackage{hyperref}
\usepackage{xr}
\usepackage{hyperref}
\makeatletter
\newcommand*{\addFileDependency}[1]{% argument=file name and extension
  \typeout{(#1)}
  \@addtofilelist{#1}
  \IfFileExists{#1}{}{\typeout{No file #1.}}
}
\makeatother
\newcommand*{\myexternaldocument}[1]{%
    \externaldocument{#1}%
    \addFileDependency{#1.tex}%
    \addFileDependency{#1.aux}%
}
\myexternaldocument{Sup-Bulk-Thin-Film-LaBa2Fe3Ox-10}
\begin{document}
\title{Pressure-induced hole delocalization in the strongly correlated quasicubic charge-transfer perovskite  \protect\ce{LaBa2Fe3O_{8+$\delta$}}}
\author{M. ElMassalami }
\affiliation{Instituto de F\'{\i}sica, Universidade Federal do Rio de Janeiro, Caixa Postal
68528, 21941-972 Rio de Janeiro RJ, Brazil}
\thanks{Dedicated to the memory of R. C. Thiel (1935--2022), R. E. Rapp (1941--2018), and M. Durieux (1923--2011).}
\author{S. Favre}
\affiliation{ Instituto de F\'{\i}sica, Facultad de Ingenier\'{\i}a, Universidad de la Rep\'ublica, Herrera y Reissig 565, CC 30, CP 11300 Montevideo, Uruguay}
\author{M. B. Silva Neto}
\affiliation{Instituto de F\'{\i}sica, Universidade Federal do Rio de Janeiro, Caixa Postal
68528, 21941-972 Rio de Janeiro RJ, Brazil}
\begin{abstract}
Analysis of the thermal and baric evolution of resistance in \ce{LaBa2Fe3O_{8+\delta}} enabled the construction of its pressure–temperature (\textit{P–T}) phase diagram, which prominently displays a critical boundary, $P^{MIT}_c(T)$, marking the transition from localized to hole-type extended states. The relatively low critical pressures [$P^{MIT}_c(T) \approx 3$–$8\,\text{GPa}$] suggest that, as $P \rightarrow P_c$ in this narrow-gap, strongly correlated charge-transfer system, both the hybridization strength and the charge-transfer character are progressively enhanced—ultimately leading to the emergence of metallicity.  Emphasizing the electronic nature of this transition, pressure-dependent structural analyses at room temperature reveal no associated structural phase transition at $P^{MIT}_c(T)$; the system retains a (weakly tetragonally distorted) quasicubic perovskite structure with Murnaghan-type compressibility up to 30\,GPa. The emergence of hole delocalization and metallic conduction, coupled with suppressed antiferromagnetism, suggests proximity to quantum criticality.
\end{abstract}
\maketitle
\section{Introduction \label{Sec-Introduction}}
Perovskite-related oxides containing 3\textit{d} transition-metal ions exhibit exceptionally rich phase diagrams, driven by the interplay of structural, electronic, and magnetic degrees of freedom, which can be tuned through controllable parameters such as pressure, temperature, magnetic fields, or chemical substitution.\cite{Imada98-MIT-Review}
This richness is exemplified by the emergence of diverse phenomena, including high-temperature superconductivity in cuprates\cite{Bednorz-and-Muller98-HTc-Cuprates,
%Lee06-Doping-Mott-Insulator-Physics-HTc-Review,Proust19-HTc-Cuprate-Ground-State-Review,Fradkin15-HTc-Interwined-Orders,
Scalapino12-UnconSCs-CommonThread} and nickelates,\cite{li19-Nickelate-SUC-Infinite-Layers,Hepting20-Electronic-structure-Nickelates} as well as colossal magnetoresistance in manganites.\cite{Tokura99-CMR-Manganites}
%Dagotto03-Magnities-CMR-Phase-Separation,Rao98-Manganate-CMR-Charge-Order,Coey95-Localization-Mixed-Valence-Manganites}
In high-temperature superconductors, particularly cuprates, control-parameter-induced evolution of phase diagrams typically encompass antiferromagnetic insulating states, strange metallic behavior, pseudogap phenomena, unconventional superconductivity,  and Fermi-liquid normal state.\cite{Scalapino12-UnconSCs-CommonThread,Lee06-Doping-Mott-Insulator-Physics-HTc-Review}
%,Proust19-HTc-Cuprate-Ground-State-Review,Fradkin15-HTc-Interwined-Orders}
Similarly, in manganites, the electronic states encompass metallic or insulating behavior, multivalent configurations, complex magnetic ordering, and phase inhomogeneities such as charge and orbital ordering.\cite{Dagotto03-Magnities-CMR-Phase-Separation,Coey95-Localization-Mixed-Valence-Manganites}
%,Rao98-Manganate-CMR-Charge-Order}

These exotic behaviors underscore the highly tunable and strongly correlated nature of 3\textit{d}-electronic systems within perovskite or related crystal structures. This observation raises an intriguing question: can other 3\textit{d}-transition metal perovskites, beyond cuprates, manganites, and nickelates, exhibit similarly rich and exotic phase diagrams?
We focus specifically on Fe-based perovskites of the form \ce{\textit{R}_{x}\textit{A}_{1-x}FeO_{3-\delta}} (\ce{\textit{A}=Ca, Sr, Ba}, \textit{R}=rare earth or Y), as some of their structural, transport, and magnetic properties are analogous to those observed in cuprates, nickelates, and manganites. At the same time, they exhibit distinctive characteristics that set them apart from these related systems\cite{Imada98-MIT-Review} (see Supplemental Material\S\ref{Sec-SM-App-Classification-Fe-Perovskite} [\Onlinecite{Comment-LaBa2Fe3Ox-Nov2025-SM}]): as an illustration, consider the phase diagram of the \ce{\textit{R}_{x}\textit{A}_{1-x}FeO_{3-\delta}}  family. Notably, it exhibits a range of transformations, including metal-insulator transition (MIT), transitions from ferromagnetic metallic to antiferromagnetic charge-transfer insulating states, and spin-state transitions between high-spin and low-spin configurations.\cite{Takano91-CaFeO3-High-to-Low-Spin-Transition,Takano77-CaFeO3-Disproportionation}
Within this family, the \ce{\textit{R}Ba2Fe3O_{8+\delta}} series\cite{91-ELmassalami91-YFe2Ba3Ox,Karen98-RBa2Fe3Ox-ND,Huang92-YBa2Fe3O8.b,GarciaGonzalez93-DyHoBa2Fe3O8,Karen03-ND-YBa2Fe3Ox,Linden99-O-order-RBa2Fe3Oz,Linden98-MES-RBa2Fe3Oz,Karen94-YBa2F3Oz-OxygenVariation,Karen05-YBa2fe3O8-structure-stoichioemetry,Linden98-LaB2Fe3Ox-MES,Gibb89-LaxBa1-xFeO3-y,96-Elzubair96-LaBa2Fe3Ox,Awana96-PrBa2Fe3O8,01-Rapp01-GdBa2Fe3Ox,99-Elzubair99-RBa2Fe3Ox-light-R,98-Elzubair98-EuBa2Fe3O,18-Camacho-La1+xBa2-xFe3O8+x} emerges as particularly relevant for our study since, in addition, they display systematic oxygen-driven reductions in unit cell volume, antiferromagnetic ordering, and electrical resistivity, making them  promising candidates for investigating novel phase transitions and electronic instabilities.

%\cite{Li92-LaBaFeOy-Phase-relation,Parras88-LaxBa1-xFeO3-y,Kharton08-LaBaFeO,Xu08-YBa2Fe3O8-E-structure,Zhang13-YBaFeO,GonzalezCalbet93-NdSmEuBa2Fe3Ox,Felner93-Y-EuBa2Fe3Ox,Elzubair99-RBaFeO,Karen98-RBa2Fe3Ox-ND,Karen03-ND-YBa2Fe3Ox,Linden99-O-order-RBa2Fe3Oz,Linden98-MES-RBa2Fe3Oz,Linden98-LaB2Fe3Ox-MES,Kharton08-LaBaFeO,Xu08-YBa2Fe3O8-E-structure,Zhang13-YBaFeO,Karen98-RBa2Fe3Ox-ND,Karen94-YBa2F3Oz-OxygenVariation,Karen05-YBa2fe3O8-structure-stoichioemetry,Li92-LaBaFeOy-Phase-relation,Elzubair99-RBaFeO,Parras88-LaxBa1-xFeO3-y}  

\ce{\textit{R}Ba2Fe3O_{8+\delta}} compounds can be broadly classified into two distinct groups based on the character of the rare-earth ion $\textit{R}^{3+}$: heavier $\textit{R}^{3+}$ ions (e.g., \ce{Dy^{3+}}, \ce{Er^{3+}}, and \ce{Y^{3+}}) and lighter $\textit{R}^{3+}$ ions (e.g., \ce{La^{3+}}, \ce{Nd^{3+}}, \ce{Sm^{3+}}, and \ce{Gd^{3+}}).
The heavier $\textit{R}^{3+}$ group
\cite{91-ELmassalami91-YFe2Ba3Ox,Karen98-RBa2Fe3Ox-ND,Huang92-YBa2Fe3O8.b,GarciaGonzalez93-DyHoBa2Fe3O8,Karen03-ND-YBa2Fe3Ox,Linden99-O-order-RBa2Fe3Oz,Linden98-MES-RBa2Fe3Oz,Karen94-YBa2F3Oz-OxygenVariation,Karen05-YBa2fe3O8-structure-stoichioemetry}
adopts a layered triple-perovskite-type structure characterized by space group P4/mmm, with ordered stacking of \ce{Ba^{2+}} and \ce{\textit{R}^{3+}} cations along the c-axis. In this configuration, Fe ions occupy octahedral and pyramidal oxygen coordination environments, resulting in predominantly trivalent Fe valency and an oxygen stoichiometry close to 8. These characteristics lead to insulating antiferromagnetic behavior, a high $T_N$, and a large magnetic moment.
In contrast, the lighter $\textit{R}^{3+}$ group 
\cite{Linden98-LaB2Fe3Ox-MES,Gibb89-LaxBa1-xFeO3-y,96-Elzubair96-LaBa2Fe3Ox,Awana96-PrBa2Fe3O8,01-Rapp01-GdBa2Fe3Ox,99-Elzubair99-RBa2Fe3Ox-light-R,98-Elzubair98-EuBa2Fe3O,18-Camacho-La1+xBa2-xFe3O8+x}
exhibits significant intermixing between \ce{\textit{R}^{3+}} and \ce{Ba^{2+}} cations, forming disordered and defect-rich perovskite structures with an average symmetry described by space group Pm$\overline{3}$m. 
This $\textit{R}^{3+}$ group exhibits non-stoichiometric oxygen content (8+$\delta$) and mixed Fe valence states (Fe$^{3+}$, Fe$^{4+}$, and Fe$^{5+}$). On the one hand, increasing the oxygen content in these compounds leads to a reduced quasicubic unit cell volume, a lower Néel temperature, diminished magnetic moments, and enhanced electrical conductivity. On the other hand, the degree of Fe valence mixing directly impacts the electronic properties, which are generally governed by the competition between the on-site Coulomb interaction and the charge-transfer energy (for further details, see \S\ref{Sec-Discussion} and \S\ref{Sec-SM-App-Classification-Fe-Perovskite} [\Onlinecite{Comment-LaBa2Fe3Ox-Nov2025-SM}]).
%\cite{18-Camacho-La1+xBa2-xFe3O8+x,Elzubair99-RBaFeO,Karen98-RBa2Fe3Ox-ND,Linden99-O-order-RBa2Fe3Oz,Linden98-LaB2Fe3Ox-MES,Kharton08-LaBaFeO,Li92-LaBaFeOy-Phase-relation,Parras88-LaxBa1-xFeO3-y} 

These remarkable features of the second class of \ce{\textit{R}Ba2Fe3O_{8+\delta}} compounds are particularly prominent in \ce{LaBa2Fe3O_{8+\delta}}.\cite{18-Camacho-La1+xBa2-xFe3O8+x,Parras88-LaxBa1-xFeO3-y, Gibb89-LaxBa1-xFeO3-y,96-Elzubair96-LaBa2Fe3Ox,Li92-LaBaFeOy-Phase-relation,Karen98-RBa2Fe3Ox-ND,Linden98-LaB2Fe3Ox-MES}
Consequently, we selected \ce{LaBa2Fe3O_{8+\delta}} as a representative system to examine how exotic its phase diagram can be and, in particular, how its charge-transfer insulating state may be driven—through external parameters such as applied pressure and oxygen loading—toward metallic behavior and, potentially, quantum criticality.

It is worth noting that \ce{LaBa2Fe3O_{8+\delta}} significantly differs from its cuprate counterpart, typically represented as \ce{La(Ba_{2-x}La_{x})Cu3O_{7-\delta}} ($0 \le x < 0.5$).\cite{Segre87-LaBa2Cu3O7-x-SUC-Oxygen-order,David87-La3Ba3Cu6O14+x-Structure,Wada89-La1+xBa2-xCu3Oy-SUC,Izumi89-La1+xBa2-xCu3O7+x-Structure} The cuprate compound crystallizes in an orthorhombic tripled perovskite cell (space group \emph{Pmmm}), featuring full occupancy of \ce{La^{3+}}  at the 1\textit{t} site and mixed \ce{Ba^{2+}} –\ce{La^{3+}}  occupancy at the 2\textit{t} site. Additionally, as $x \rightarrow 0$ and $\delta \rightarrow 0$, it manifests metallicity and ultimately becomes superconducting with $T_c \approx 93\,K$. A variable-range hopping (VRH) conduction mechanism (see Eq.\ref{Eq.VRH-Resistivity-T-dependent}) has also been observed in insulating, tetragonally-tripled \ce{La(Ba_{2-x}La_{x})Cu3O_{7-\delta}} for $x \ge 0.5$, characterized by an activated energy of $T_o \approx 23\,K$.\cite{Segre87-LaBa2Cu3O7-x-SUC-Oxygen-order}  We show below that: (i) the value of $T_o$ for \ce{LaBa2Fe3O_{8+\delta}} is six orders of magnitude higher, even higher than what is observed for semiconductors,\cite{Shklovskii84-Electronic-Prop-DopSemiconduct} highlighting how far the Fe-based system lies from metallic behavior; and (ii) applying pressure results in a strong reduction of $T_o$ in the Fe-based system.

The pronounced dependence of \ce{LaBa2Fe3O_{8+\delta}} properties on oxygen content stems from variations in the O(2p) electronic bands. Clearly, hybridization between highly correlated Fe(3d) states and the more band-like O(2p) levels can be further tuned by applying physical or chemical pressure. In this study, we performed pressure-dependent resistivity measurements ($2 \leq T \leq 400\,K$, $P \leq 2.8\,GPa$, $H \leq 140\,kOe$) to construct a \textit{P–T} phase diagram for \ce{LaBa2Fe3O_{8+\delta}}. Our primary findings highlight the critical role of pressure in promoting hole delocalization and broadening the oxygen-hole band, effectively reducing the ratio of the charge-disproportionation gap to the oxygen-hole bandwidth ($\Delta_{CD}/W$).\cite{Kawasaki98-Fe(d4)-pervoskite-Phase-Transitions}
%
%%%%%%%%%%%%%%%%%%%%%%%%%%%%%%%%%%%%%%%%%%%%%%%%%%%%%%%%%%%%%%%%%
\section{Experimental \label{Sec-Experimental}}
We prepared a series of bulk and thin-film samples of \ce{LaBa2Fe3O_{8+\delta}}; the bulk ones were prepared following procedures detailed in Ref.\,\Onlinecite{18-Camacho-La1+xBa2-xFe3O8+x}, while thin films were grown using pulsed laser deposition technique (see \S\ref{Sec-SM-Thin-Film-Synthesis} [\Onlinecite{Comment-LaBa2Fe3Ox-Nov2025-SM}]).

Ambient-pressure room-temperature X-ray diffraction (XRD) patterns for all samples were recorded using Cu($K_{\alpha}$) radiation.
For high-pressure measurements, room-temperature diffraction data were collected using a cryogenic diamond anvil cell at the XDS beamline (LNLS, Brazil), with $\lambda = 0.6199$\,\AA. 
The pressure-transmitting medium (PTM) used in this study is helium gas.  
Given that, even above its solidification pressure ($\sim$12.1\,GPa), this medium remains an effective pressure-transmitting medium,\cite{Klotz09-Hydrostatic-limits-P-Transmitting-Media} the challenging discrepancies encountered during Rietveld refinement are not attributed to non-hydrostaticity related to this medium. Rather, the discrepancies most likely arise  from the combined effects of (i) the structural complexity of the solid solution—featuring La$^{3+}$/Ba$^{2+}$ cation intermixing and oxygen non-stoichiometry—and (ii) extrinsic non-hydrostatic conditions, likely caused by overfilling the pressure chamber.
Pressure was monitored using the ruby fluorescence method.

Le Bail pattern-matching and Rietveld refinements were carried out for all diffraction patterns using the \texttt{FullProf} Suite software package.\cite{Rodriguez-Carvajal-FullProf} In the Rietveld analysis, atomic positions and site occupancies were fixed, while thermal parameters were adopted from Ref.\,\Onlinecite{Karen98-RBa2Fe3Ox-ND}.

Oxygen content in both bulk and thin-film samples was estimated from a calibration curve relating the oxygen content to the \textit{a}-parameter of the quasicubic perovskite cell, as detailed in Ref.\,\Onlinecite{Karen98-RBa2Fe3Ox-ND} (see \S\ref{Sec-SM-Oxygen-Count-Aging-Effect} [\Onlinecite{Comment-LaBa2Fe3Ox-Nov2025-SM}] for additional remarks).

Magnetization and DC susceptibility measurements were conducted using a Quantum Design\!\texttrademark{}  SQUID magnetometer ($|H|\le 70\,kOe, 2 \le T \le 300\,K$).
Electrical resistances, $R(T,P,H)$, were measured using a conventional four-terminal technique within a  PPMS (Quantum Design\!\texttrademark) environment $(2\le T\le 400\,K, H\le 140\,kOe,10\,nA \le I \le 10\,mA)$. 
For high-temperature resistivity data, refer to Ref.\,\Onlinecite{18-Camacho-La1+xBa2-xFe3O8+x}. 
Pressure-dependent resistances were measured using a piston-type pressure cell (up to $P \leq 2.8$\,GPa) with Daphne 7373 oil as the pressure-transmitting medium and lead (Pb) as the pressure gauge.
$R(T,P,H)$ measurements primarily targeted the VRH regime (100\,$< T <$\,300\,K) since, above this range, the resistance follows Arrhenius-type conduction, whereas below it—extending down to liquid helium temperatures—it exceeds the instrument's measurement limits due to insulating antiferromagnetic behavior.

Due to sample-specific characteristics—such as irregular geometry—absolute resistivity could not be reliably determined. Van der Pauw-type measurements under ambient conditions yielded a range of $0.2 \lesssim \rho \lesssim 2\,\Omega\!\cdot\!\text{cm}$, spanned by oxygen non-stoichiometry: This establishes the bounds for the true $\rho(T, P, H)$.
%
%%%%%%%%%%%%%%% begin Figure 1 %%%%%%%%%%%%%%%%%%%%%%%trim={<left> <lower> <right> <upper>}
\begin{figure}[hptb!] 
\centering
\noindent
\includegraphics[scale=0.29,trim={0.0cm 0cm 0.0cm 0cm},clip]{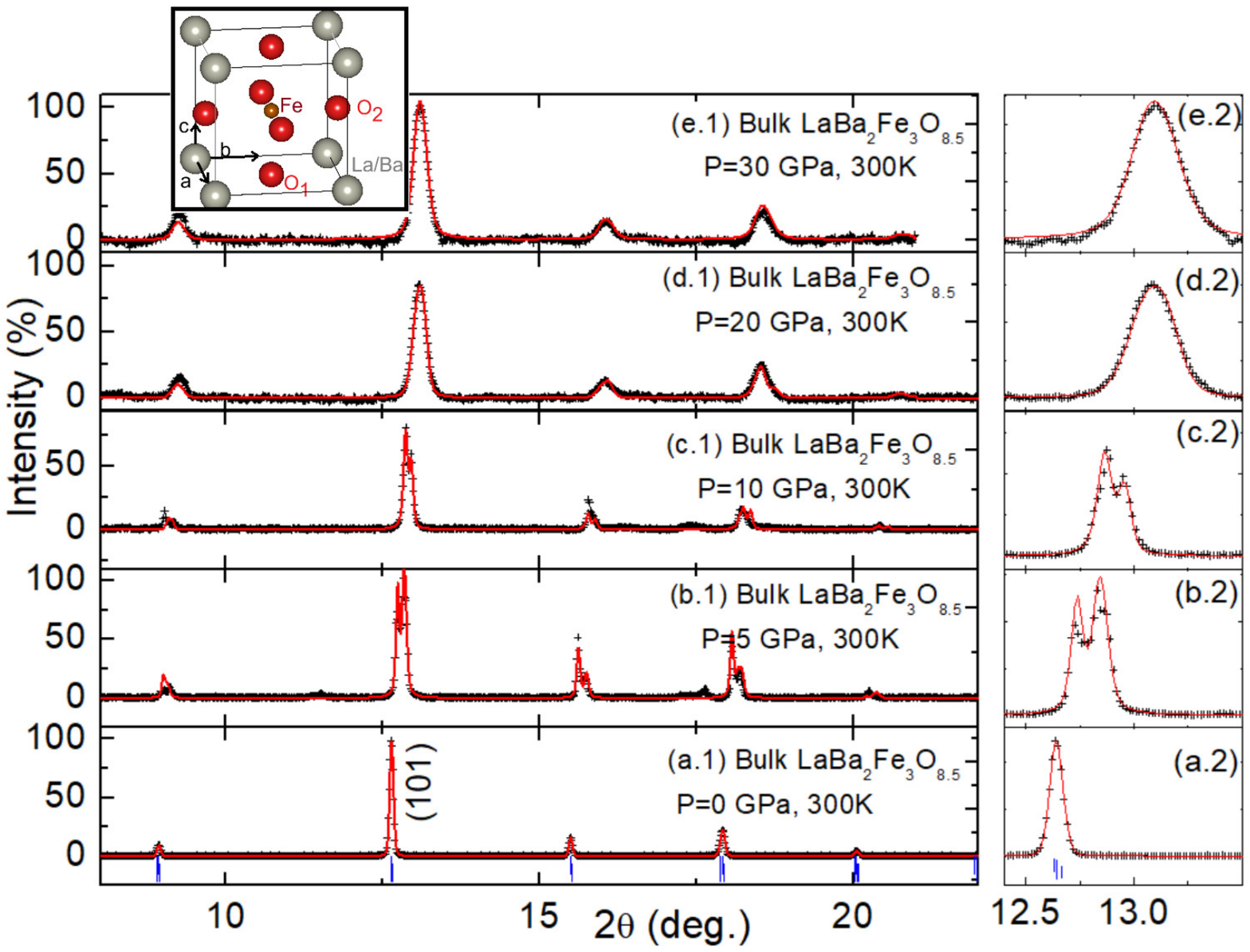}%
\caption{Baric evolution of room-temperature diffractograms of \protect\ce{LaBa2Fe3O_{8.5}} measured using synchrotron radiation.  
\textbf{(a.1)} $P = 0$\,GPa.  
\textbf{(b.1)} $P = 5.0$\,GPa.  
\textbf{(c.1)} $P = 10.0$\,GPa.  
\textbf{(d.1)} $P = 20.0$\,GPa.  
\textbf{(e.1)} $P = 30.0$\,GPa.  
Panels \textbf{(a.2)}–\textbf{(e.2)} show expanded views of the intense (101) peak from the corresponding panels \textbf{(a.1)}–\textbf{(e.1)}, highlighting its pressure-induced evolution.\cite{Note-LaBa2Fe3Ox-2025-Fig1-Impurity-Peaks}  
The \texttt{+} symbols, red continuous lines, and short blue bars represent the measured intensities, calculated patterns, and Bragg reflection positions, respectively.
{\textit{Inset}}: Schematic representation of the crystallographic unit cell.
}
\label{Fig1-LaBa2Fe3Ox-Diffractigrams-Bulk}%
\end{figure}
%%%%%%%%%%%%%%% end Figure 1 %%%%%%%%%%%%%%%%%%%%%%%
%
%%%%%%%%%%%%%%% begin Figure 2 %%%%%%%%%%%%%%%%%%%%%%%trim={<left> <lower> <right> <upper>}
\begin{figure}[hptb!] 
\centering
\noindent
\includegraphics[scale=0.27,trim={0.0cm 0cm 0.0cm 0cm},clip]{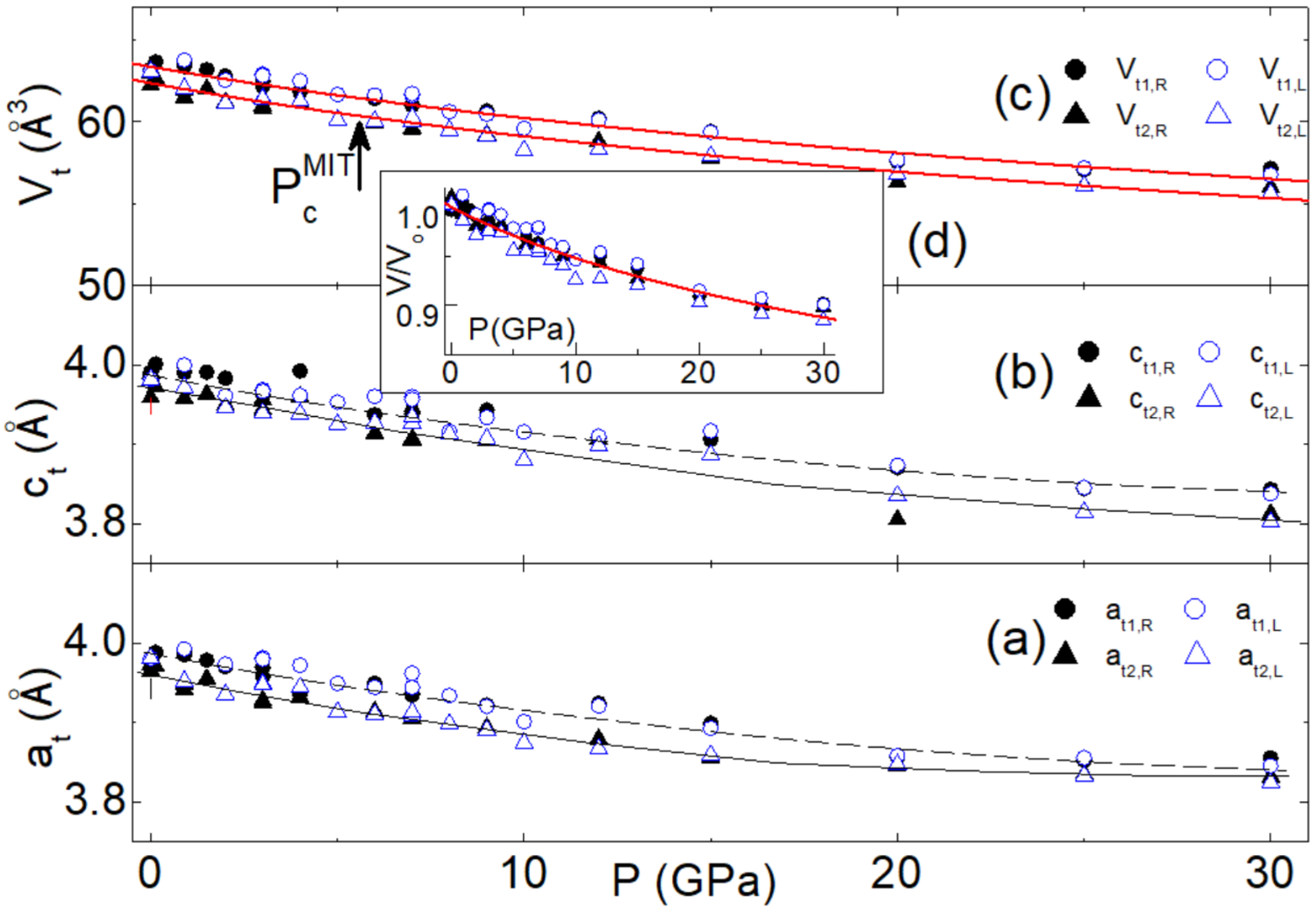}%
\caption{
Baric evolution of the lattice parameters of bulk \ce{LaBa2Fe3O_{8.5}}, based on the structural model  described in the text.  Parameters  extracted using Rietveld $R$ (Le Bail $L$) refinement are denoted by filled (empty) symbols.  Circles (triangles)  correspond to the first $t1$ (second $t2$)  tetragonal cell.
\textbf{(a)} \text{a}-parameters. 
\textbf{(b)}  \text{c}-parameters.  
\textbf{(c)} Unit-cell volumes. Vertical arrow indicates the absence of any structural phase transition at $P^{\mathrm{MIT}}_c(300\,\text{K}) = 5.6(6)$\,GPa (see Fig.\ref{Fig5-P-T-Phase-Diagram}).  
\textbf{(d)} Normalized $V/V_0$. Thin curves serve as visual guides, while the solid red curve represents a fit to Eq.\ref{Eq.V-vs-P-Murnaghan}.
For all parameters, error bars are within the size of plotted symbols.
}
\label{Fig2-LaBa2Fe3Ox-Bulk-Lattice-P-vary}%
\end{figure}
%%%%%%%%%%%%%%% end Figure 2 %%%%%%%%%%%%%%%%%%%%%%%
%
\section{Results and Analysis \label{Sec-Results-Analysis}}
\subsection{Baric Evolution of Crystalline Structure \label{SubSec-Structure}}
Analysis of ambient-pressure structural properties of both bulk and thin-film \ce{LaBa2Fe3O_{8+\delta}} confirms a stable quasicubic perovskite structure, with no indication of unit-cell doubling or tripling (see \S\ref{Sec-SM-Structure-Ambient-Pressure} [\Onlinecite{Comment-LaBa2Fe3Ox-Nov2025-SM}]). This quasicubic framework remains stable down to 10\,K. However, noticeable line broadening and shoulder features\cite{Karen98-RBa2Fe3Ox-ND} near the prominent (101) peak suggest the coexistence of two weakly tetragonally distorted phases. These distortions are attributed to \ce{La^{3+}/Ba^{2+}} intermixing and oxygen out-diffusion, which result in distinct surface and bulk oxygen stoichiometries. In the following, we examine the effects of pressure on the structural properties of bulk \ce{LaBa2Fe3O_{8+\delta}} sample.

The pressure-dependent evolution of the diffractograms for bulk \ce{LaBa2Fe3O_{8.5}} is presented in Fig.\ref{Fig1-LaBa2Fe3Ox-Diffractigrams-Bulk}. 
Notably, the (101) peak, detailed in the right-hand panels, undergoes a distinct shift towards higher angles accompanied by nonmonotonic changes in line shape. 
The baric evolution of these diffraction were analyzed assuming a minimal model comprising two basic tetragonal perovskite unit cells  (see \S\ref{Sec-SM-Structure-Ambient-Pressure} [\Onlinecite{Comment-LaBa2Fe3Ox-Nov2025-SM}]). Moreover, we consider that at low pressures ($P < 5$\,GPa), the positions of Bragg peaks from both cells are closely spaced, justifying the commonly adopted quasicubic approximation. With increasing pressure to the intermediate range ($5 < P < 10$\,GPa), relative shifts between the peaks of the two cells become more pronounced, leading to visibly resolved peak splitting.
At higher pressures, however, non-hydrostatic conditions dominate, causing a reduction in compressibility, flattening of the lattice parameters, and broadening of the Bragg peaks—resulting in the appearance of highly broadened and possibly overlapping features in the diffraction patterns.

 Based on this minimal model, we analyzed the diffractograms using both Le Bail pattern-matching and Rietveld refinement methods. Representative Rietveld refinements are presented in Fig.\ref{Fig1-LaBa2Fe3Ox-Diffractigrams-Bulk}. Furthermore, the refined lattice parameters as functions of pressure—shown in Fig.\ref{Fig2-LaBa2Fe3Ox-Bulk-Lattice-P-vary}—are in good agreement between the two approaches, within experimental uncertainties. 
This consistency provides confidence that the baric evolution depicted in Fig.\ref{Fig2-LaBa2Fe3Ox-Bulk-Lattice-P-vary}, aside from deviations caused by non-hydrostatic conditions at high pressure, reliably reflects the actual baric evolution of the system.

Figure~\ref{Fig2-LaBa2Fe3Ox-Bulk-Lattice-P-vary} indicates that both tetragonal cells exhibit the same trend: a monotonic volume reduction with increasing pressure, without any indication of structural transitions or anomalous intermediate behavior.
Indeed, the pressure-dependent volume variations of both unit cells collapse on each other, as illustrated in Fig.\ref{Fig2-LaBa2Fe3Ox-Bulk-Lattice-P-vary}(d). Furthermore, as evident in Figs.\ref{Fig2-LaBa2Fe3Ox-Bulk-Lattice-P-vary}(c), this monotonic volume contraction closely matches predictions based on the Murnaghan equation:
\begin{equation}
V(P)=V_{0}\left[1+P\left({\frac  {B^\prime_{0}}{B_{0}}}\right)\right]^{-1/B^\prime_0},
\label{Eq.V-vs-P-Murnaghan}
\end{equation}
where $V_{0}$, $B_{0}$, and $B'_{0}$ denote, respectively, the ambient-pressure volume, the bulk modulus, and the pressure derivative of the bulk modulus evaluated at $P=0$\,GPa.
Fitting the empirical $ V(P) $ to Eq.\ref{Eq.V-vs-P-Murnaghan}, as shown in Figs.\ref{Fig2-LaBa2Fe3Ox-Bulk-Lattice-P-vary}(c,d), yields the following:  
For the first tetragonal cell, $ V_{0,t_1} = 63.36\,\text{\AA}^3 $, $ B_{0,t_1} = 160\,\text{GPa} $, and $ B'_{0,t_1} = 8.0 $; for the second cell, $ V_{0,t_2} = 62.36\,\text{\AA}^3 $, $ B_{0,t_2} = 150\,\text{GPa} $, and $ B'_{0,t_2} = 8.0 $.  
Both ambient-pressure volumes $ V_{0,t_1} $ and $ V_{0,t_2} $ (surface and interior regions, respectively) lie within the reported ambient-pressure volume range of 61.32–63.56\,\AA$^3$, corresponding to variations in oxygen non-stoichiometry.\cite{Karen98-RBa2Fe3Ox-ND} Moreover, the bulk moduli $ B_{0} $ obtained for both cells agree closely with reported values for analogous Fe-based perovskites; for instance, $ B_{0} \approx 146\,\text{GPa} $ for \ce{SrFeO3}.\cite{Kawazoe23-SrFeO3-Bulk-Modulus,kawakami03-SrFeO3-High-Press-MES-XRD}
Finally, it is worth noting that the differential stress arising from non-hydrostatic conditions likely influences the extracted bulk modulus parameters. In addition, this stress accounts for the apparent lattice stiffening and anisotropic compression observed in Fig.\ref{Fig2-LaBa2Fe3Ox-Bulk-Lattice-P-vary}, as well as the peak broadening seen in Fig.\ref{Fig1-LaBa2Fe3Ox-Diffractigrams-Bulk}(d--e).
%
%%%%%%%%%%%%%%% begin Figure 3 %%%%%%%%%%%%%%%%%%%%%%%trim={<left> <lower> <right> <upper>}
\begin{figure}[hptb!] 
\centering
\noindent
\includegraphics[scale=0.27,trim={0.0cm 0cm 0.0cm 0cm},clip]{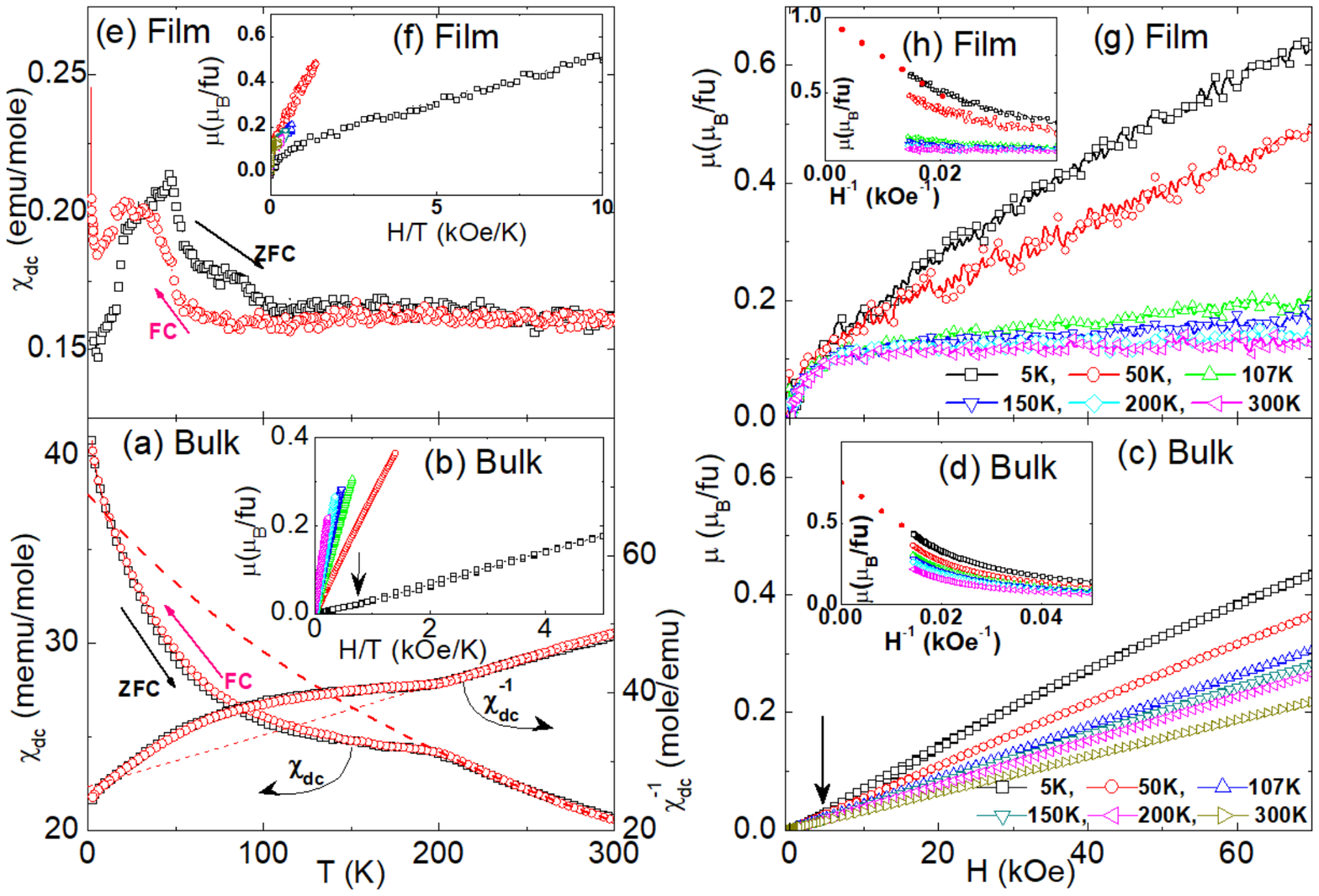}%
\caption{
Ambient-pressure magnetic properties of bulk $\ce{LaBa2Fe3O_{8.75}}$ and thin-film \ce{LaBa2Fe3O_{8.70}}.  
\textbf{(a)} $\chi_{dc}(1\,\text{kOe}, T)$ and $\chi^{-1}_{dc}(1\,\text{kOe}, T)$ of bulk sample. Dashed line represents $\frac{13.35}{T + 352}$ emu/mole.  
\textbf{(b)} $\mu^{fu}(T,H)$ versus $H/T$ of  powdered sample.  Short vertical arrows indicate a weak but noticeable spin-flop feature, a characteristic of a AFM-ordered state.
\textbf{(c)} $\mu^{fu}(T,H)$ of  powdered sample.  
\textbf{(d)} $\mu^{fu}$ versus $\nicefrac{1}{H}$ of powdered sample. 
\textbf{(e)} $\chi_{dc}(1\,\text{kOe}, T)$ of thin-film.\cite{Note-LaBa2Fe3Ox-2025-No-Inverse-Susc-Films} 
\textbf{(f)} $\mu^{fu}(T,H)$ versus $H/T$ of thin-film.  
\textbf{(g)} $\mu^{fu}(T,H)$ of  thin-film.  
\textbf{(h)} $\mu^{fu}$ versus  $\nicefrac{1}{H}$ of thin-film. Dotted lines  are linear extrapolation to $\lim_{\nicefrac{1}{H} \to 0} \mu^{fu}(H)$.
}
\label{Fig3-Powder-ThinFilm-MvsH-XvsT}
\end{figure}
%%%%%%%%%%%%%%% end Figure 3 %%%%%%%%%%%%%%%%%%%%%%%%
\subsection{Ambient-Pressure Magnetization \label{SubSec-Magnetization}}
The ambient-pressure magnetic properties of bulk and thin-film $\ce{LaBa2Fe3O_{8+\delta}}$ are presented in Fig.\ref{Fig3-Powder-ThinFilm-MvsH-XvsT}. The thermal evolution of the dc susceptibility, $\chi_{dc}(1\,\text{kOe}, T)$, for the powdered bulk sample, shown in Fig.\ref{Fig3-Powder-ThinFilm-MvsH-XvsT}(a), exhibits Curie-Weiss behavior with an effective moment of $\mu^{fu}_{eff} = 3.4\,\mu_B$ and a Curie-Weiss (CW) temperature $\theta_{CW} = 352\,\text{K}$. A significant deviation from the Curie-Weiss behavior occurs below a magnetic event at 200 K. However, at temperatures well below this 200K-event, another Curie-Weiss-type behavior emerges, indicating that while this event involves partial co-orientation of magnetic moments, a major portion remains paramagnetic down to 2 K. 
It is possible that these two portions are related to the two tetragonally-distorted phases corresponding to surface and interior regions.
Two key observations provide further insight into the nature of the 200\,K event: 
(i)  Mossbauer spectroscopy\cite{Gibb89-LaxBa1-xFeO3-y,96-Elzubair96-LaBa2Fe3Ox,Linden98-LaB2Fe3Ox-MES} reveals that,   around 200\,K, the magnetic moments undergo a slow relaxation process. (ii) The plot of $\mu^{fu}(T,H)$ versus $H/T$ in Fig.\ref{Fig3-Powder-ThinFilm-MvsH-XvsT}(b) shows a collapse of all curves at $T \geq 107\,\text{K}$, indicating that the 200\,K event does not lead to a long-range magnetic order across the whole sample; $T_N$ is expected to emerge only below 100\,K.
Additionally, the isothermal $\mu^{fu}(5\,\text{K},H)$ curve, shown in Fig.\ref{Fig3-Powder-ThinFilm-MvsH-XvsT}(c), reinforces the non-saturated character of the magnetization.  

The magnetic properties of thin-film samples, shown in Figs.\ref{Fig3-Powder-ThinFilm-MvsH-XvsT}(e–h), exhibit several similarities and some distinctions, compared to those of the powdered sample discussed above. 
$\chi_{dc}(1\,\text{kOe}, T)$, Fig.\ref{Fig3-Powder-ThinFilm-MvsH-XvsT}(e), reveals that the warming and cooling curves—zero-field-cooled (ZFC) and field-cooled (FC), respectively—coincide above approximately 100 K, as also observed in the bulk case. 
Below 100 K, however, a marked ZFC–FC irreversibility develops (which was only weakly present in the powder case).
A maximum in $\chi_{dc}(1\,\text{kOe}, T)$ is manifested near 40 K.
Above this maximum, $\mu^{fu}(T,H)$ \textemdash plotted as a function of $H/T$ in Fig.\ref{Fig3-Powder-ThinFilm-MvsH-XvsT}(f) \textemdash collapse onto a single curve, again indicating the absence of total  long-range order above this maximum. 

The saturation moment, $\mu_{\mathrm{sat}}^{\mathrm{Fe}}$, was obtained by extrapolating the $\mu^{\mathrm{fu}}(5\,\text{K},H)$ magnetization data shown in Figs.\ref{Fig3-Powder-ThinFilm-MvsH-XvsT}(d,h). For bulk samples, $\mu_{\mathrm{sat}}^{\mathrm{Fe}} = \lim_{1/H \to 0} \mu_{\mathrm{sat}}^{\mathrm{fu}}(H)/3 \approx 0.25\,\mu_B$, while for thin films, $\mu_{\mathrm{sat}}^{\mathrm{Fe}} = \lim_{1/H \to 0} \mu_{\mathrm{sat}}^{\mathrm{fu}}(H)/3 \approx 0.33\,\mu_B$. These values clearly indicate a substantial reduction in the Fe magnetic moment in \ce{LaBa2Fe3O_{8+\delta}}.

The magnetism of bulk samples is consistent with previous studies.\cite{Gibb89-LaxBa1-xFeO3-y,96-Elzubair96-LaBa2Fe3Ox,Linden98-LaB2Fe3Ox-MES,Karen98-RBa2Fe3Ox-ND} In contrast, the magnetic behavior of the thin-film samples is best understood by decomposing the isothermal $\mu^{\mathrm{fu}}(T,H)$ curves, as in e.g. Fig.\ref{Fig3-Powder-ThinFilm-MvsH-XvsT}(g), into two distinct contributions:
(i) a soft ferromagnetic component (most probably of extrinsic origin, related to the weak peaks observed in XRD analysis); and
(ii) an antiferromagnetic component, with saturation occurring only at high fields, originating from the ordering of magnetic moments that are paramagnetic above 40\,K.
The FM contribution is clearly manifested across the entire measured temperature range, with $\mu^{\mathrm{fu}}_{\mathrm{sat}}$ spanning approximately $0.7$–$1\,\mu_B$/fu. This persistent FM component explains the absence of a clear CW behavior in thin-films: the moment induced by the CW contribution is smaller than that of the FM background. It also accounts for the higher $\mu^{\mathrm{fu}}_{\mathrm{sat}}$ of thin films—by approximately $0.1 \sim 0.3\,\mu_B$/fu—compared to bulk samples.

These results collectively suggest that La$^{3+}$/Ba$^{2+}$ cation intermixing and oxygen non-stoichiometry in \ce{LaBa2Fe3O_{8+\delta}} strongly suppresses magnetic moments and ordering. This is evidenced by: (i) the significantly reduced ordering temperature, 
(ii) the measured effective moment $\mu^{exp}_{eff}(Fe) \approx 3.4\,\mu_B$ being lower than the theoretically calculated  spin-only high-spin $\mu^{\mathrm{cal}}_{\mathrm{eff}}(Fe^{3+})\approx 5.9\,\mu_B$ or $\mu^{\mathrm{cal}}_{\mathrm{eff}}(Fe^{4+}) \approx 4.9\,\mu_B$, and 
(iii) the measured $\mu_{sat}^{exp} (Fe)\approx 0.3\,\mu_B$ being drastically lower than the calculated spin-only high-spin  $\mu^{\mathrm{cal}}_{\mathrm{sat}}(Fe^{3+}) \approx 5\,\mu_B$ or $\mu^{\mathrm{cal}}_{\mathrm{sat}}(Fe^{4+})\approx4\,\mu_B$. 
%%%%%%%%%%%%%%% begin Figure  4 %%%%%%%%%%%%%%%%%%%%%%%trim={<left> <lower> <right> <upper>}
\begin{figure}[hptb!] 
\centering
\noindent
\includegraphics[scale=0.27,trim={0.0cm 0cm 0.0cm 0cm},clip]{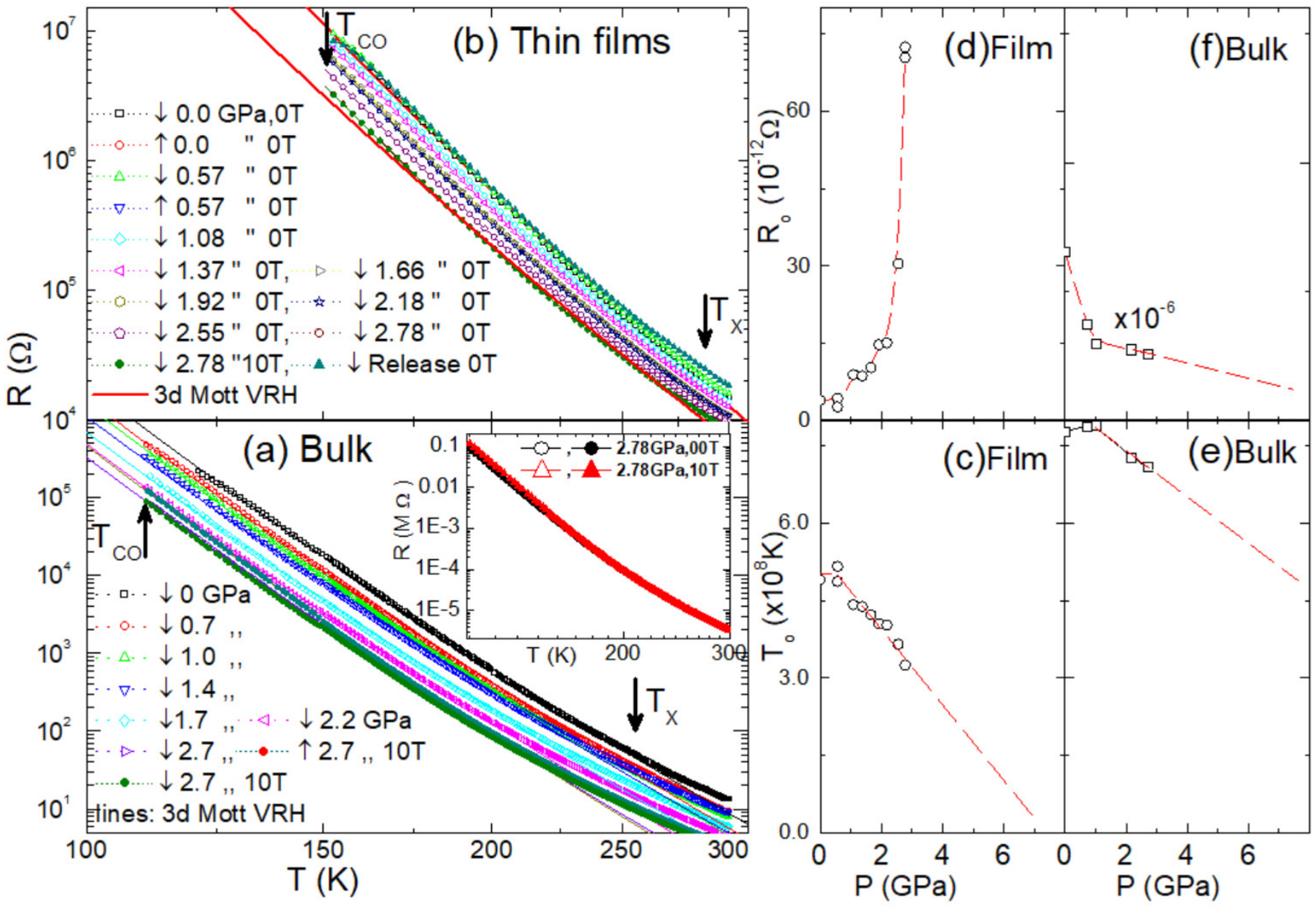}%
\caption{ 
\textbf{(a)} $R(P,T,H)$ of bulk \ce{LaBa2Fe3O_{8.75}} in a log-log format.
\textbf{(b)} $R(P,T,H)$ of thin-film \ce{LaBa2Fe3O_{8.70}} in a log-log format. 
Downward (upward) arrows in the legend indicate measurements during cooling (warming). 
Open symbols,  filled circles, filled  triangles correspond, respectively, to measurements taken during pressure increase,
at $P = 2.78\,\mathrm{GPa}$ and $10\,\mathrm{T}$, and after returning to ambient pressure.    
Fitting $R(P,T,H)$ to Eq.\ref{Eq.VRH-Resistivity-T-dependent}  give \textbf{(c)} $T_o(P)$ and \textbf{(d)}$R_o(P)$ of thin films while \textbf{(e)} $T_o(P)$ and \textbf{(f)} $R_o(P)$ of bulk samples.
Dashed lines in panels (c-f) are guides to the eye, while solid lines in panels (a-b) represent VRH fits.
\textit{Inset of panel (a)}: $R(P=2.78\,\text{GPa}, T, H=0\,\text{T})$ shown in circles and $R(P=2.78\,\text{GPa}, T, H=10\,\text{T})$ in triangles.
}
\label{Fig4-LaBa2Fe3Ox-RpvsT-ThinFilm}%
\end{figure}
%%%%%%%%%%%%%%% end Figure 4 %%%%%%%%%%%%%%%%%%%%%%%%

%%%%%%%%%%%%%%%%%
\subsection{Baric Evolution of Resistance \label{Sec-Resistivity-Baric-Evolution}}
Ambient-pressure resistivity of \ce{LaBa2Fe3O_{8+\delta}} have been extensively studied (see \S \ref{Sec-SM-Resistivity-Ambient-Pressure} [\Onlinecite{Comment-LaBa2Fe3Ox-Nov2025-SM}]). The main feature of its thermal evolution can be classified into three distinct temperature regimes, separated by two events, the charge disproportionation at  $ T_{CD}$ and charge order at $T_{CO} $: 
(i) The high-temperature metallic state ($T > T_{CD}$) attributed to itinerant $e_g^{1\pm\nu}$ electrons with homogeneous Fe valence ($\ce{Fe}^{4\pm\nu}$); 
(ii) The intermediate-temperature range ($T_{CO} < T < T_{CD}$) characterized by thermally activated transport, attributed to thermally-activated hole hopping. Here, the resistivity manifests an absence of dependence on an applied magnetic field (at least for |H|$\le$ 140 kOe and P$\le$2.8GPa) indicating a decoupling between magnetic and charge transport degrees of freedom. 
Finally, (iii) the low-temperature regime ($T < T_{CO} $) where the system becomes insulating, antiferromagnetic, and charge ordered.  Below, we analyze the pressure-dependent influence on the thermal evolution of the resistivity within the intermediate regime.

\begin{figure}[hptb!] 
\centering
\noindent
\includegraphics[scale=0.27,trim={0.0cm 0.0cm 0.0cm 0.0cm},clip]{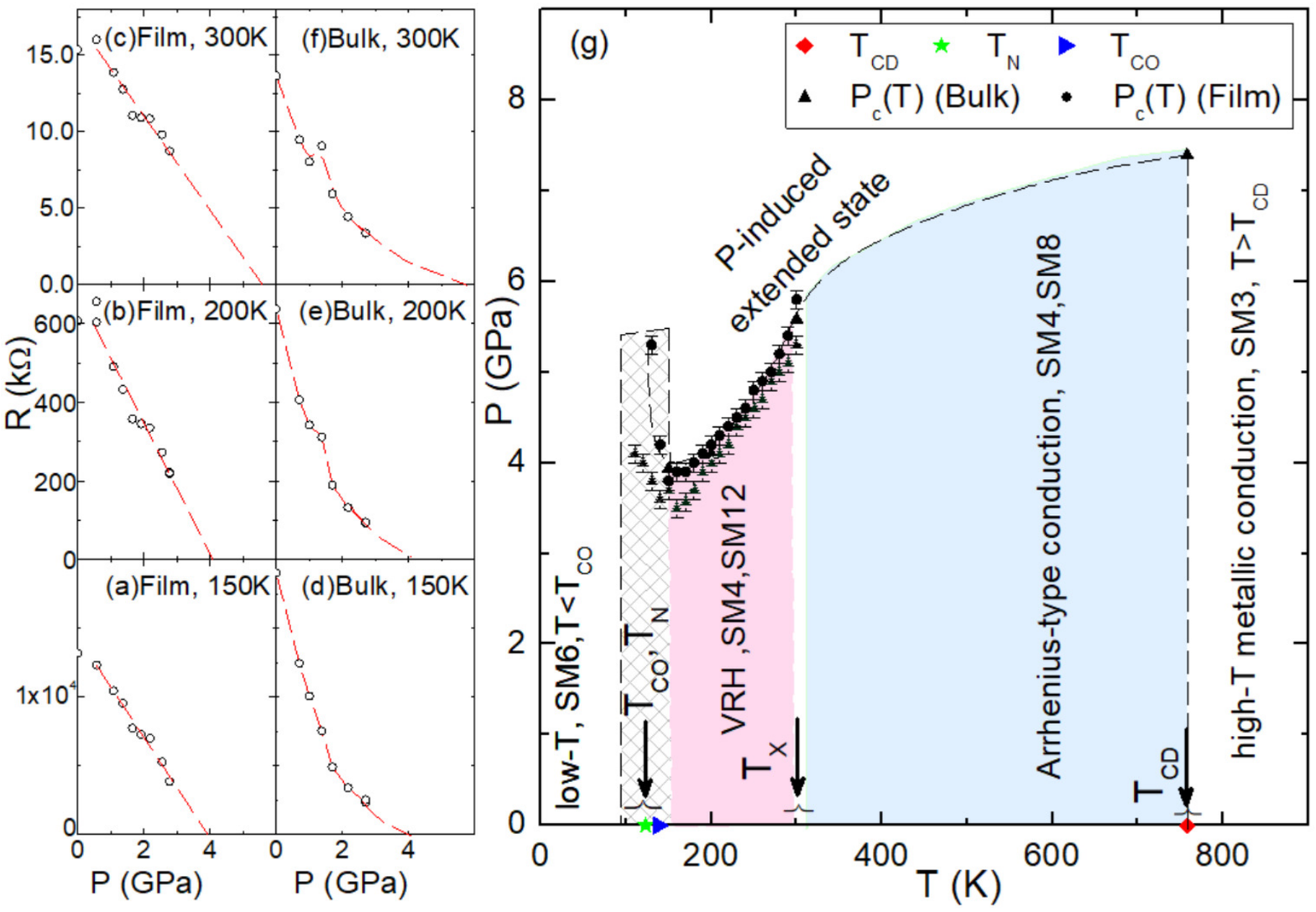}%
\caption{
Representative $R(P,T_f=150,200,300\,K)$ of thin-films \textbf{(a–c)} and bulk \textbf{(d–f)},  used to determine the critical pressures by extrapolating resistance to zero.  
\textbf{(g)} \textit{P–T} phase diagram (see main text and \S\ref{Sec-SM-App-Classification-Fe-Perovskite} [\protect\Onlinecite{Comment-LaBa2Fe3Ox-Nov2025-SM}]). 
The hatched region indicates the temperature range over which $T_{CO}$ and $T_N$ are distributed; its upper range is constrained by the available pressure range. 
The intermediate $T_{CO}\le T \le T_{CD}$  range is characterized by thermally activated transport, comprising both Arrhenius-type and variable-range hopping (VRH) conduction regimes (see \S\ref{Sec-SM-App-Classification-Fe-Perovskite} and \S\ref{Sec-SM-Resistivity-Ambient-Pressure} 
[\protect\Onlinecite{Comment-LaBa2Fe3Ox-Nov2025-SM}]); $T_X$ 
denote the crossover temperature. 
The low $T \le T_{CO}$ regime is classified as insulating, antiferromagnetic, and charge ordered.
}
\label{Fig5-P-T-Phase-Diagram}% 
\end{figure}

%A charge-ordering transition,  at $T_{CO}$ event would be manifested as an upward resistivity jump  $T_{CO}$) is found to be strongly pressure-dependent but largely unaffected by H = 10 T. 
Figures~\ref{Fig4-LaBa2Fe3Ox-RpvsT-ThinFilm}(a,b) present the pressure-dependent resistance, $R(P,T)$, of bulk and thin-film samples. 
The data are shown within $T_{CO} < T < 300$\,K range, where  the resistance follows a variable-range hopping (VRH) behavior; for $T > T_{X} \approx$300\,K, it crosses over to an Arrhenius-type conduction (see \S \ref{Sec-SM-Resistivity-Ambient-Pressure} [\Onlinecite{Comment-LaBa2Fe3Ox-Nov2025-SM}]). Below $T_{CO}$, charge order emerges. 
The most notable effect of pressure is the monotonic decrease in resistance across the studied intermediate temperature range. 
This baric reduction in $R(P,T)$ is further illustrated by the pressure evolution of representative isothermal resistances $R(P,T_f)$ at $T_f = 150, 200$, and $300\,\text{K}$, as shown in Figs.\ref{Fig5-P-T-Phase-Diagram}(a–f). 
These isotherms—and similar ones—decrease monotonically with increasing pressure.
Extrapolating each curve to zero resistance at fixed temperatures yields the corresponding critical pressures, $P_c(R_{T_f} \rightarrow 0)$, which signify the onset of an extended electronic state regime where charge excitation and hopping become increasingly facile.
The thermal evolution of these critical pressures is plotted in Fig.\ref{Fig5-P-T-Phase-Diagram}(g): as temperature decreases downwards to $T_{CO}$, $P_c(R_{T_f} \rightarrow 0)$ is monotonically decreased.  Below, $T_{CO}$ a sharp increase in $P_c(R_{T_f} \rightarrow 0)$ is observed—signaling the onset of charge ordering and indicating that higher pressures are required to induce this MIT transition.
It is worth adding the following two points: {\bf (i)} the resulting phase diagram in Fig.\ref{Fig5-P-T-Phase-Diagram}(g) incorporates all $P_c(R_{T_f} \rightarrow 0)$ values, along with key ambient-pressure transition temperatures ($T_{CD}$, $T_N$, and $T_{CO}$), under the assumption that $P_c(T_{CD}) = P_c^{MIT}(T_o = 0\,K)$ (note the sample-dependent distribution of these parameters). {\bf (ii)}  Although the phase diagram identifies a transition at $P^{MIT}_{c}(300\,K)=5.6(6)$ GPa, the room-temperature Murnaghan plot (Fig.\ref{Fig2-LaBa2Fe3Ox-Bulk-Lattice-P-vary}) does not reveal a distinct structural transformation at this critical pressure within current experimental resolution.  The absence of an observable pressure-induced structural transition is consistent with similar observation for \ce{LaSr2Fe3O9} up to approximately 50 GPa.\cite{Nasu02-HP-MES-Fe-Oxides,Kawakami02-LaSrFe3O9-P-induced-FM-Charge-Uniform} 

All $R(P,T)$ curves of Fig.\ref{Fig4-LaBa2Fe3Ox-RpvsT-ThinFilm}(a,b)  were analyzed using the usual expression for Mott´s variable range hopping resistivity (see the derivation at the end of \S \ref{Sec-SM-App-Classification-Fe-Perovskite} [\Onlinecite{Comment-LaBa2Fe3Ox-Nov2025-SM}]):
\begin{eqnarray}
    \rho_{VRH}(T)&=&\rho^0_{VRH}\exp\left[\left(\nicefrac{T_o}{T}\right)^{1/4}\right],\nonumber\\
    \rho^0_{VRH}&=&1/e^2 N(E_F)\nu_{ph}R_{Mott}^2,\nonumber\\
    T_o&=&\frac{4\nu_c}{k_B N(E_F)\xi^3},
    \label{Eq.VRH-Resistivity-T-dependent}
\end{eqnarray}
where $R_{Mott}$ is an optimal hopping distance, $\nu_c$ is a geometric constant, $\nu_{ph}$ is a measure of the (constant) density of phonon states, and $\xi$ is the localization length—the characteristic spatial scale over which an electronic wavefunction remains appreciable before decaying exponentially (see \S \ref{Sec-SM-App-Classification-Fe-Perovskite} [\Onlinecite{Comment-LaBa2Fe3Ox-Nov2025-SM}]). 

The fit of the above-mentioned VRH model of Eq.\ref{Eq.VRH-Resistivity-T-dependent} to the experimental $R(P\le 2.78\,GPa,T_{CO} \le T \le T_X)$ are illustrated by the solid lines in Fig.\ref{Fig4-LaBa2Fe3Ox-RpvsT-ThinFilm}(a,b).
The corresponding fit parameters are shown in Figs.\ref{Fig4-LaBa2Fe3Ox-RpvsT-ThinFilm}(c-f). 
 
Just like the magnetic properties,  there are similarity and difference among the baric evolution of fit parameters of bulk and thin-film samples.  As an example, while $T_o$  of both bulk and thin-film are decreasing with pressure, the rate of suppression of $T_o $ of thin-film is much stronger. 
Specifically,  Fig.\ref{Fig4-LaBa2Fe3Ox-RpvsT-ThinFilm}(c) indicates that $ T_o $ of thin-film remains nearly constant up to approximately 0.5\,GPa; beyond that, it exhibits a sharp and monotonic decline, with a linear regression yielding a slope of  $\frac{\partial T_o}{\partial P} \approx -7.3 \times 10^7\,\text{K/GPa}$.  
Extrapolating $ T_o(P) $ to zero gives the critical pressure  $P_c^{'} = P_{c}({T_o \rightarrow 0}) \approx 7.4(6)\,\text{GPa}$. 

The combination of the analytical expression in Eq.\,\ref{Eq.VRH-Resistivity-T-dependent} with the empirical pressure dependences of $\rho^{0}_{\mathrm{VRH}}(P)$, $T_{0}(P)$, and $R(P,T_f)$ shown in Figs.\ref{Fig4-LaBa2Fe3Ox-RpvsT-ThinFilm}–\ref{Fig5-P-T-Phase-Diagram} allows a qualitative assessment of the baric evolution of $N(E_F)(P)$, $\xi(P)$, $\nu_{\mathrm{ph}}(P)$, and $R_{\mathrm{Mott}}(P)$ (or their products). Substituting $\rho^{0}_{\mathrm{VRH}}$ and $T_{0}$ into Eq.\,\ref{Eq.VRH-Resistivity-T-dependent} at fixed temperature $T_f$ yields the pressure–dependent isothermal resistance
\begin{equation}
R(P,T_f)\;\propto\;
\frac{\exp\!\left\{\left[\dfrac{\mathrm{constant}}{N(E_F)\,\xi^{3}}\right]^{1/4}\right\}}
{N(E_F)\,\nu_{\mathrm{ph}}\,R_{\mathrm{Mott}}^{2}}\,,
\label{Eq.Baric-VRH-R(P,T_f)-LaBaFeO}
\end{equation}
where all parameters on the right-hand side are pressure dependent, except for the constant. 
Comparing Eq.\,\ref{Eq.Baric-VRH-R(P,T_f)-LaBaFeO} with experiment leads to the following inferences.
\textbf{(i)} Because $T_{0}\!\propto\![N(E_F)\,\xi^{3}]^{-1}$, the observed decrease of $T_{0}(P)$ as $P\!\to\!P_c^{\mathrm{MIT}}$ implies that $N(E_F)\,\xi^{3}$ \emph{increases} with pressure. This trend is naturally attributed to a substantial growth of $\xi$ (with at most a modest increase in $N(E_F)$). An analogous pressure-driven delocalization—enhanced $\xi$ accompanied by reduced $T_{0}$—has been reported in elemental Te.\cite{Oliveira21-Pressure-Induced-Anderson-Mott-Tellurium}
\textbf{(ii)} Since both $R(P,T_f)$ and $T_{0}(P)$ decrease with pressure [Fig.\ref{Fig5-P-T-Phase-Diagram}(a–f)], the exponential factor in the numerator of Eq.\,\ref{Eq.Baric-VRH-R(P,T_f)-LaBaFeO} must diminish sufficiently to outweigh any denominator changes; i.e., the pressure-enhanced $N(E_F)\,\xi^{3}$ reduces the exponent and drives $R(P,T_f)$ downward. The observed increase of $R_{0}(P)$ in thin films versus its decrease in bulk [Figs.\ref{Fig4-LaBa2Fe3Ox-RpvsT-ThinFilm}(d,f)] is then attributed to different pressure dependences of the product $\nu_{\mathrm{ph}}\,R_{\mathrm{Mott}}$ (and implicitly of $\xi$ through $R_{\mathrm{Mott}}$), reflecting material-specific effects (e.g., disorder, microstructure, and strain) in films versus bulk.
 %
%%%%%%%%%%%%%%% begin Figure 5 %%%%%%%%%%%%%%%%%%%%%%%trim={<left> <lower> <right> <upper>}
%%%%%%%%%%%%%%% end Figure 5 %%%%%%%%%%%%%%%%%%%%%%%
\section{Discussion and Conclusions \label{Sec-Discussion}}
It is evident that oxygen non-stoichiometry and $A$-site intermixing of \ce{La^{3+}} and \ce{Ba^{2+}} jointly render Fe in \ce{LaBa2Fe3O_{8+\delta}} intrinsically mixed valent (see \S\,\ref{Sec-SM-App-Classification-Fe-Perovskite} [\Onlinecite{Comment-LaBa2Fe3Ox-Nov2025-SM}]).
This multivalency is well supported by M\"ossbauer,\cite{Linden98-LaB2Fe3Ox-MES,96-Elzubair96-LaBa2Fe3Ox} magnetization,\cite{18-Camacho-La1+xBa2-xFe3O8+x,96-Elzubair96-LaBa2Fe3Ox} and neutron diffraction\cite{Karen98-RBa2Fe3Ox-ND} studies. It also follows directly from the formula
$\ce{La^{3+}_{1}Ba^{2+}_{2}(Fe^{(4-\nu)}_{1-\beta}Fe^{(4+\nu)}_{\beta})_{3}O^{2-}_{8+\delta}}$,
which, under charge neutrality (Eq.\,\ref{Eq.Multi-valency-Neutrality}), yields
$\nu = (3 - 2\delta)/(3 - 6\beta)$, implying coexisting \ce{Fe^{4-\nu}} and \ce{Fe^{4+\nu}} cations and explicitly reflecting the dependence on oxygen non-stoichiometry and charge redistribution.

The effect of oxygen incorporation/removal on Fe valence is conveniently expressed using Kr\"oger--Vink notation.\cite{Linden98-LaB2Fe3Ox-MES} Oxygen incorporation increases hole concentration and drives Fe toward higher oxidation state:
$O_{2} + 2V_{O}^{\bullet\bullet} \rightleftarrows 4h^{\bullet} + 2O_{O}^{\times},$
whereas oxygen loss yields vacancies and electrons,
$2O_{O}^{\times} \rightleftarrows 4e^{'} + O_{2} + 2V_{O}^{\bullet\bullet},$
thereby reducing Fe valence. 
Although the Fe valence is mostly set by the oxygen content (non-stoichiometry $\delta$), the resulting Fe valence states are \emph{dynamic}, depending on $T$ and $P$. Variations in $T$ and $P$ modify the charge-transfer gap, hybridization, and defect equilibria (oxygen vacancy/incorporation), thereby shifting the balance among Fe valence states and altering the transport, magnetic response, and pressure-driven MIT characteristics of \ce{LaBa2Fe3O_{8+\delta}}.

The most prominent feature of the \textit{P–T} phase diagram of \ce{LaBa2Fe3O_{8+\delta}} is the characteristic boundary  $P_c(R_{T_f} \rightarrow 0)$, which delineates the transition from localized to extended electronic states without destabilizing or distorting the crystal structure.  Similar phase boundaries also appear in the diagrams of the related perovskites \ce{SrFeO3}, \ce{CaFeO3}, and \ce{LaSr2FeO9} (\S\ref{Sec-SM-App-Classification-Fe-Perovskite} [\Onlinecite{Comment-LaBa2Fe3Ox-Nov2025-SM}]) and is usually attributed to a reduction in the charge-transfer gap and a broadening of the band in which the electrons related to the $e_g(Fe)-p_\sigma$ are delocalized.\cite{Nasu02-HP-MES-Fe-Oxides,Kawakami02-LaSrFe3O9-P-induced-FM-Charge-Uniform,kawakami03-SrFeO3-High-Press-MES-XRD,Kawakami05-SrFeO3-CaFeO3-Sr2LaFe3O9} 
Along the same framework, the empirical analysis at the end of \S\ref{Sec-Resistivity-Baric-Evolution} shows that applied pressure increases $\xi$ and $N(E_F)$. Together, these effects promote a crossover from variable-range hopping toward extended electronic states.

The critical pressures observed across the entire \textit{P–T} phase diagram of \ce{LaBa2Fe3O_{8+\delta}} are significantly lower than those reported for other Fe-based perovskites (Fig.\ref{FigSM1-P-T-Diagram-Perovskites} [\Onlinecite{Comment-LaBa2Fe3Ox-Nov2025-SM}]). 
This  indicates a stronger pressure-driven reduction of an already small charge-transfer gap, together with a more pronounced pressure-induced widening of an already broad electronic bandwidth.
Several factors likely contribute: (i) below $T_{CD}$, mixed-valence Fe facilitates intersite hopping, effectively reducing the charge-transfer energy; and (ii) a higher average Fe valence with enhanced Fe–O hybridization broadens both the Fe~$3d$ and O~$2p$ bands, further narrowing the gap. In a charge-transfer framework, the relevant control parameter is $\Delta/W$; thus, a reduction of this ratio lowers the threshold pressure for transition toward extended electronic states.
 It is worth emphasizing that, although metal–insulator transitions are widely observed in 3\textit{d} transition-metal perovskite oxides,\cite{Imada98-MIT-Review} this pressure-induced MIT in \ce{LaBa2Fe3O_{8+\delta}} is particularly notable, as it emerges in a regime where the antiferromagnetic state is already strongly suppressed (see \S\ref{SubSec-Magnetization} and related discussion).
 
Our study demonstrates that both the thin-film format and its bulk counterpart provide favorable conditions for suppressing the antiferromagnetic state and lowering the critical pressure required to access the extended-state regime. Ongoing investigations aim to disentangle the respective contributions of cationic disorder, oxygen non-stoichiometry, sample morphology, and synthesis route in driving these behaviors. In parallel, we are extending the pressure range to higher values to further explore the electronic properties and potential instabilities that may emerge at or beyond the extended-state transition. 
Another area of interest concerns the inference \textemdash{} based on a  close analysis of the phase diagram of \ce{LaBa2Fe3O_{8+\delta}} (Fig.\ref{Fig5-P-T-Phase-Diagram}) and those of related Fe-based systems (Fig.\ref{FigSM1-P-T-Diagram-Perovskites} [\Onlinecite{Comment-LaBa2Fe3Ox-Nov2025-SM}]) \textemdash{} that  there is  possible presence of quantum criticality, whose influence extends into the adjacent finite-temperature regime. Within this quantum critical region, quantum fluctuations are expected to dominate over thermal ones, potentially leading to non-Fermi-liquid behavior in both thermodynamic and transport properties. These fluctuations may also favor the emergence of exotic phases. Further analysis is underway.

In summary, the pressure-dependent structural analysis of \ce{LaBa2Fe3O_{8+\delta}} indicates that its quasicubic perovskite structure remains stable \textemdash within our current experimental resolution \textemdash  up to pressures as high as 30\,GPa. Specifically, while weak tetragonal distortions are observed, no structural phase transitions occur at room temperature throughout the investigated pressure range; only Murnaghan-type volumetric compressions are detected.\\
Transport studies, however, demonstrate distinct critical behaviors associated with hole dynamics under variations in temperature and pressure. The characteristic temperatures and pressures defining these critical events are summarized in the \textit{P–T} phase diagram of Fig.\ref{Fig5-P-T-Phase-Diagram}. Analysis of these critical events and their corresponding phase boundaries underscores the uniqueness of \ce{LaBa2Fe3O_{8+\delta}} compared to other related charge-transfer-type, \ce{Fe^{4+}}-bearing perovskites such as $\ce{\textit{A}_{1-x}\textit{A}^{$\prime$}_{x}FeO_{3-\delta}}$ and  $\ce{\textit{R}_{x}\textit{A}_{1-x}FeO_{3-\delta}}$ (\ce{\textit{A,A}^{$\prime$}=Ca,Sr,Ba}, \textit{R}=La). Specifically, \ce{LaBa2Fe3O_{8+\delta}} exhibits a smaller electronic gap and stronger charge-transfer interactions. Consequently, even moderate applied pressures significantly enhance the spatial extent of the charge-carrier wavefunctions, facilitating the delocalization of holes above a critical pressure threshold.\\
Another notable difference between \ce{LaBa2Fe3O_{8+\delta}} and these related Fe-based perovskites is that the pressure-induced hole delocalization is emerging within a reduced antiferromagnetic state. These findings position \ce{LaBa2Fe3O_{8+\delta}} as a promising system for exploring emergent phenomena, particularly quantum criticality, through the controlled manipulation of parameters such as applied pressure, chemical doping, cationic disorder, and substrate-induced strain.
%.
%%%%%%%%%%%%%%%%%%%%%%%%%%%%%%%%%%%%%%%%%%%%%%%%%%%%
\begin{acknowledgments}
We acknowledge the interest and collaborative support of Dr. R. D. dos Reis and Dr. J. C. Tenorio. High-pressure structural measurements were carried out at the Brazilian Synchrotron Light Laboratory (LNLS), with the assistance of the XDS beamline staff [Proposal \#20160107]. S.F. acknowledges partial financial support from the Uruguayan PEDECIBA program and the Sistema Nacional de Investigadores–ANII.
\end{acknowledgments}
\bibliography{Tex-Bulk-Thin-Film-LaBa2Fe3Ox-10.bbl}

\begin{thebibliography}{67}
\bibitem [{Not({\natexlab{a}})}]{Note-LaBa2Fe3Ox-2025-Fig1-Impurity-Peaks}%
  \BibitemOpen
  \href@noop {} {} ({\natexlab{a}}),\ \bibinfo {note} {{Few }extremely weak
  extrinsic peaks—e.g., at $d \approx 2.02$ and $3.09\,\text{\AA}$ in
  Fig.1(b.1,c.1)—are non-systematic and are not indicative of P-induced
  structural changes. As such, they do not influence the analysis or
  conclusions of this work and were excluded from the Rietveld and Le Bail
  analyses}\BibitemShut {NoStop}%
\bibitem [{Not({\natexlab{b}})}]{Note-LaBa2Fe3Ox-2025-No-Inverse-Susc-Films}%
  \BibitemOpen
  \href@noop {} {} ({\natexlab{b}}),\ \bibinfo {note} {{Unlike} bulk case in
  Fig.3(a), the inverse susceptibilities are not shown, as no CW behavior is
  observed at higher temperatures.}\BibitemShut {Stop}%
\end{thebibliography}
\end{document}